# Prompt Engineering for Responsible Generative AI Use in African Education: A Report from a Three-Day Training Series

*Organized By:*

*Generative AI for Education and Research in Africa (GenAI-ERA)*

**Participating Authors and Committee**

Benjamin Quarshie[1], Vanessa Willemse[2], Macharious Nabang[3],
Bismark Nyaaba Akanzire[4], Patrick Kyeremeh[5], Saeed Maigari[6],
Dorcas Adomina[7], Ellen Kwarteng[8], Eric Kojo Majialuwe[9],
Craig Gibbs[10], Jerry Etornam Kudaya[11], Sechaba Koma[12],
Matthew Nyaaba[13]

[1] Department of Creative Arts, Mampong Technical College of Education, Asante Mampong, Ghana
[2] Northern Cape Provincial Department of Education, South Africa
[3] Department of Creative Arts, Bagabaga College of Education, Tamale, Ghana
[4] Department of Education, Gambaga College of Education, Gambaga, Ghana
[5] Department of Mathematics and ICT, St. Joseph's College of Education, Ghana
[6] Department of Physical Planning and Municipal Services, Federal University of Transportation, Daura, Nigeria
[7] Department of English Language Education, University of Education, Winneba, Ghana
[8] Department of Arts and Social Sciences, St. Joseph's College of Education, Ghana
[9] University of Manitoba, Canada / University for Development Studies, Ghana
[10] JET Education Services, Johannesburg, South Africa
[11] Generative AI for Education and Research in Africa (GenAI-ERA)
[12] Department of Science Education, National University of Lesotho, Lesotho
[13] Department of Educational Theory and Practice, University of Georgia, USA

**Corresponding Author: matthew.nyaaba@uga.edu**

**Abstract**

Generative artificial intelligence (GenAI) tools are increasingly used in education, yet many educators lack structured guidance on responsible prompt engineering, particularly in African and other resource-constrained contexts. This case report documents a three-day professional development training series organised by Generative AI for Education and Research in Africa (GenAI-ERA), aimed at building educators' and researchers' capacity to use prompt engineering ethically and effectively for academic writing, teaching, and research. The programme was delivered online and involved 468 participants from multiple African countries, including university educators, postgraduate students, and researchers. The training followed a progressive structure, moving from foundational prompt design to applied and advanced strategies, including ethical prompting and persona-guided interactions. Data sources included programme registration surveys (n = 468), webinar interaction records, facilitator observations, and transcripts of the three sessions. These data were analysed using descriptive statistics and computationally supported qualitative techniques, including thematic indicators, reflexivity markers, and sentiment analysis. Key insights indicate that participants increasingly recognised prompt engineering as a form of AI literacy that requires ethical awareness, contextual sensitivity, and pedagogical judgement rather than technical skill alone. The case also highlights persistent practical needs in low-resource settings, including access to affordable AI tools, locally relevant training materials, and sustained institutional support. This case report contributes practical evidence on how structured, scaffolded prompt-engineering training can support responsible GenAI use in digital education and academic writing, particularly within African higher education contexts. The case also reveals persistent gaps in access to structured AI training, locally relevant materials, and institutional guidance. Based on these insights, the case recommends sustained professional development in prompt engineering across all levels of education, from basic to tertiary, and the formal integration of prompt literacy into curricula and inclusive AI initiatives in Africa

Contents

# 1. Purpose of the Report

Generative Artificial Intelligence (AI) technologies such as ChatGPT, Gemini, and Copilot are transforming how knowledge is produced, communicated, and consumed (Ibarra-Sáiz et al., 2025). While these tools enable automation, creativity, and wider access to information, they also pose complex ethical, pedagogical, and epistemic challenges, especially for educators and researchers in the Global South (Nyaaba & Zhai, 2024). In many African contexts, disparities in technological infrastructure and digital literacy limit equitable participation in the emerging AI economy.

Recognizing this gap, Generative AI for Education and Research in Africa (GenAI-ERA), a non-profit, Pan-African organization, launched a three-day professional-development series entitled *"Empowering Educators and Researchers through Responsible and Innovative Prompt Engineering"(See Appendix A).* The initiative sought to build foundational competence in communicating with AI systems through prompt engineering and to cultivate ethical and culturally responsive use of generative technologies. This report presents an analytical account of what transpired across the three sessions, synthesizing participant interactions, emergent questions, and pedagogical outcomes. It highlights how prompt engineering can become a vehicle for advancing AI literacy, critical reflection, and research productivity across African educational institutions.

# 2. Prompt Engineering in Education

Recent scholarship increasingly positions prompt engineering as a critical interface between human intent and generative AI behaviour, particularly in educational contexts. Studies such as *Integrating Visual Context into Language Models for Situated Social Conversation Starters* demonstrate that well-designed prompts can anchor AI outputs in contextual, social, and situational cues, improving relevance and interpretability in learning interactions (Janssens et al., 2024). Similarly, *Prompt Engineering Using ChatGPT: Crafting Effective Interactions and Building GPT Apps* highlights how structured prompts that specify role, task, constraints, and output format significantly enhance system reliability and pedagogical usefulness (Tabatabaian, 2024). In education, this reinforces the idea that prompts are not neutral instructions but epistemic tools that shape how knowledge is represented, scaffolded, and evaluated by AI systems.

Within instructional settings, prompt engineering has been shown to support curriculum-aligned content generation and reflective teaching practices. Research on *Automatic Lesson Plan Generation via Large Language Models with Self-Critique* illustrates how iterative prompting and self-evaluation mechanisms can improve lesson coherence, instructional sequencing, and alignment with learning objectives (Zheng et al., 2024). Likewise, *Learning to Prompt in the Classroom to Understand AI Limits* demonstrates that engaging teachers and students in prompt design deepens their understanding of AI affordances and limitations, fostering critical engagement rather than passive consumption. These studies suggest that prompt engineering functions as a form of instructional design literacy, requiring educators to articulate pedagogical goals clearly, anticipate learner needs, and critically assess AI-generated outputs.

From an AI literacy perspective, prompt engineering is increasingly recognised as both a technical and ethical skill. Emerging frameworks emphasise that effective prompting requires awareness of bias, data provenance, and contextual sensitivity, particularly in culturally diverse and resource-constrained environments (Nyaaba & Zhai, 2025). Instructional prompt strategies such as role prompting, constraint-based prompting, example-based prompting, and reflective prompting enable educators to guide AI outputs while maintaining human oversight and pedagogical agency. In the Global South, where access to advanced AI infrastructure and institutional guidance remains uneven, prompt engineering offers a transferable, low-cost pathway to meaningful AI integration (Murungu, 2024). By foregrounding prompt literacy as a core component of AI literacy, educational systems can empower educators not only to use generative AI tools effectively, but also to interrogate their limitations, align them with local curricula, and deploy them responsibly in teaching, learning, and research.

## 3. Approach

This report presents a programme evaluation of the GenAI-ERA three-day webinar series on prompt engineering. The evaluation adopted a qualitative, reflexivity-oriented approach to examine how participants engaged with the training content, the kinds of questions and misconceptions that emerged, and how discussions evolved across sessions. The approach was suitable for an organisation seeking to document learning processes and strengthen future professional development designs.

### 3.2 Setting, Programme Structure, and Participants

The training was delivered as a three-session (lasting approximately 90 minutes for each section with variant extension for plenary) webinar series held from **July 3 to July 5, 2025**. Each session was designed using a progressive learning sequence, moving from foundational concepts to applied prompting strategies and advanced techniques. Participants included educators, postgraduate students, and researchers across multiple African countries. Participant role, interests, AI experience level, and tool usage were captured through the programme participation form (n = 468).

### 3.3 Data Sources

Three data sources informed the report:

1. *Programme participation form (n = 468):* Used to describe participants' professional roles, interest areas, prior experience with AI tools, and AI tools previously used.

2. *Webinar interaction data:* Facilitator notes and observations captured recurring questions, misconceptions, and moments of conceptual clarification during live sessions.

3. *Webinar transcripts:* Session recordings were transcribed and used as the primary text corpus for analysing thematic emphasis, reflexivity markers, and sentiment patterns across sessions.

## 3.4 Data Preparation

Transcripts were prepared for analysis through standard text-processing steps, including removal of non-informative symbols, basic cleaning, and segmentation into analysable text units. Where needed, the dataset was processed using Python in Google Colab to support systematic handling of transcripts.

## 3.5 Analytic Strategy

To examine patterns of engagement and emphasis across the three webinar sessions, the analysis combined qualitative interpretive reading with descriptive, computationally supported text analytics. The analytic strategy was exploratory and indicator-based, designed to surface trends in discussion focus, confidence expressed, and affective orientations rather than to only measure learning outcomes or causal effects. To wit, four complementary analytic layers were applied to the webinar transcripts, namely:

- **Progressive learning indicators:**
  Selected learning-oriented terms (for example, *learn, understand, experience*) were tracked across transcript segments to identify shifts in discussion emphasis and confidence across sessions. These indicators were used descriptively to examine how participants' language evolved over time, rather than as direct measures of cognitive development or knowledge acquisition.
- **Reflexivity indicators:**
  Self-referential linguistic markers (for example, *I think, I feel, we discussed*) were examined to surface patterns of expressed personal positioning, judgement, and confidence during discussions. These markers functioned as descriptive signals of reflexive expression, not as indicators of reflective depth or critical consciousness.
- **Thematic analysis:**
  Predefined pedagogical and technology-related themes were operationalised through clustered keywords to identify dominant areas of discussion across sessions. This approach enabled a structured overview of topical emphasis while retaining interpretive flexibility during qualitative reading of transcript excerpts.
- **Sentiment analysis:**
  The VADER sentiment analysis approach was applied to transcript segments to estimate whether discussion tone was predominantly positive, neutral, or negative. Sentiment scores were interpreted as general affective orientations within the learning environment rather than precise emotional states.

All transcript preprocessing (including cleaning, tokenisation, stop-word removal, and lemmatisation) and visualisation were conducted using Python in Google Colab to support systematic handling of the dataset (Similar to approach espoused by Quarshie et al., 2025). Findings were presented through descriptive interpretation supported by visual outputs such as bar charts, line plots, and heatmaps. Given the reliance on keyword-based and lexicon-driven approaches, results are interpreted as indicative patterns of participation and discussion dynamics, not as definitive evidence of learning impact.

### 3.6 Ethical Considerations

The report draws on programme data for evaluation and learning-improvement purposes. No sensitive personal information was reported. Participation in the surveys was voluntary, and responses were analysed in aggregate for reporting and programme improvement purposes, with no personally identifiable information disclosed. Participant identities were not disclosed, and quotes, where used, were presented without names or identifying details. Where session recordings were used, the analysis relied on publicly available or consented materials associated with the webinar series.

### 3.7 Registering Participants

Registration for the webinar was open and voluntary, and participants enrolled through an online Google Forms registration form. The registration form also served as a pre-webinar survey designed to capture participants' professional background, areas of interest, prior experience with AI tools, and expectations for the workshop. This pre-survey enabled the facilitators to identify participants' learning needs, readiness levels, and priority areas, which directly informed the pacing, examples, and emphasis of the workshop sessions. The pre-webinar survey collected both biodata (such as professional role and institutional context) and needs-based information, including interest in research use, classroom teaching, ethical AI, and prompt engineering. Participants were able to select multiple options where applicable, allowing for a nuanced understanding of overlapping roles and interests.

Following the completion of the three-day workshop, participants were invited to complete a post-webinar feedback survey, also administered through Google Forms. This follow-up instrument gathered participants' reflections on the workshop content, perceived learning gains, relevance to their professional practice, and overall satisfaction. The feedback responses were used to assess the effectiveness of the workshop, identify areas of conceptual clarification, and inform recommendations for future AI-literacy and prompt-engineering training initiatives.

*3.7.1 Participants and professional roles*

Participants represented a diverse professional mix across education and research (see Figure 1). The largest group comprised tertiary/university educators (n = 187; 40.0%), followed by postgraduate students (n = 162; 34.6%), researchers (n = 84; 17.9%), and undergraduate students (n = 83; 17.7%). Additional participants included primary/secondary educators (n = 54; 11.5%),

AI practitioners (n = 16; 3.4%), and respondents who selected other roles (n = 41; 8.8%), with 14 (3.0%) specifying their roles under the "If other" option. (*NB: Participants could select more than one role; therefore, category totals exceed 468*)

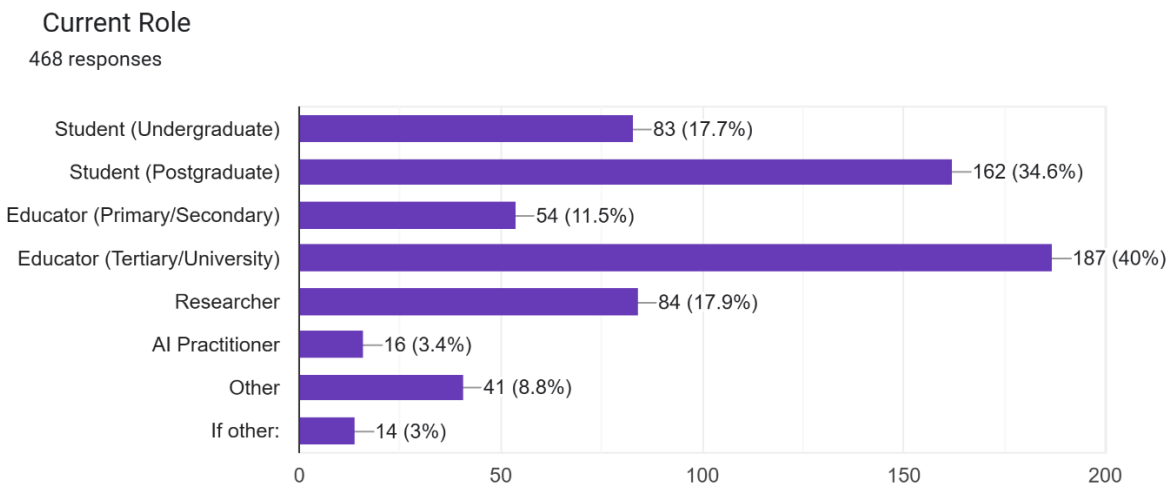

*Figure 1: Participants and professional roles*

.

### 3.7.2 Participant Interest Areas and Training Relevance

A total of 468 respondents indicated the topics they were most interested in pursuing through the GenAI-ERA training series (see Figure 2). Responses show strong demand for practical, education-facing use of generative AI, with a clear emphasis on research productivity, classroom teaching, and student learning. The most selected topic was AI for Research and Academic Work (n = 399; 85.3%), suggesting that participants were highly motivated to improve academic writing workflows, literature engagement, synthesis, and research support practices using generative AI tools. This was followed by AI for Classroom Teaching (n = 340; 72.6%) and AI for Student Learning (n = 298; 63.7%), indicating a strong interest in translating AI knowledge into everyday instructional practice and learner support.

Interest in responsible and skills-based AI use was also substantial. Ethical AI Use was selected by 287 respondents (61.3%), signalling that many participants approached AI adoption with an awareness of integrity, bias, and responsible application. Relatedly, participants expressed considerable interest in skill development through Advanced Prompting Techniques (n = 277; 59.2%) and Building AI-Powered Educational Tools (n = 263; 56.2%), reflecting a readiness to move beyond basic awareness into higher-level prompt design and applied innovation. While Basic Prompt Engineering still attracted meaningful interest (n = 226; 48.3%), the pattern suggests that many participants were seeking more than introductory knowledge; they wanted immediately applicable strategies for teaching, research, and tool creation.

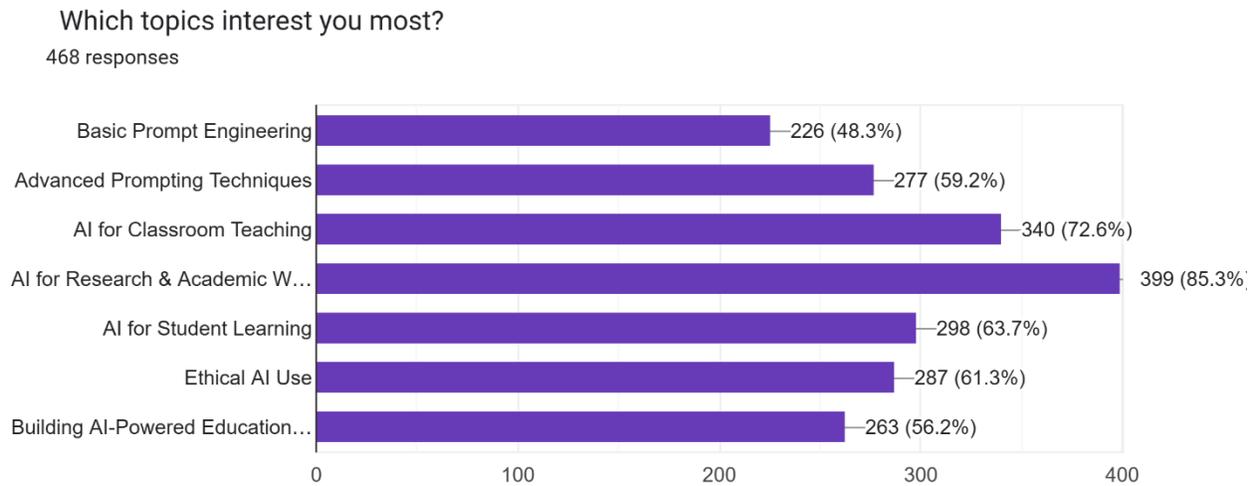

*Figure 2: Participant Interest Areas*

### 3.7.3 Participants' Experience with AI Tools and Usage Patterns

An analysis of participants' prior experience with AI tools revealed a clear readiness gap that directly informed the design and necessity of the webinar series (see Figure 3). Although most respondents had some level of exposure to generative AI, the majority were positioned at the early-to-mid stages of adoption. Specifically, 40.4% identified as beginners who had only experimented with AI tools a few times, while 40.0% reported intermediate use on an occasional basis. Only a small proportion described themselves as advanced users (8.8%) or experts who train others, with 10.0% indicating no prior experience at all. This profile suggests widespread interest in AI alongside limited depth, confidence, and strategic competence.

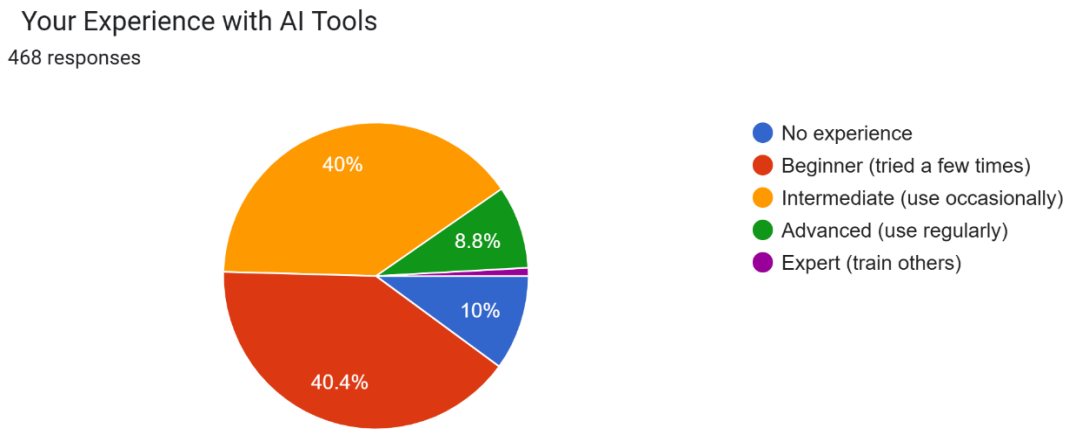

Figure 3: *Participants' prior experience with AI tools*

Patterns of AI tool usage further clarified the nature of this gap. ChatGPT emerged as the dominant entry point into generative AI, used by 87.2% of participants, while engagement with other platforms such as Microsoft Copilot, Gemini/Bard, and Claude was comparatively limited. A notable minority had not used any AI tools prior to the programme (see Figure 4). These trends point to a reliance on easily accessible, text-based systems and limited exposure to broader AI ecosystems, reinforcing the need for guidance that extends beyond tool familiarity to principled, transferable skills.

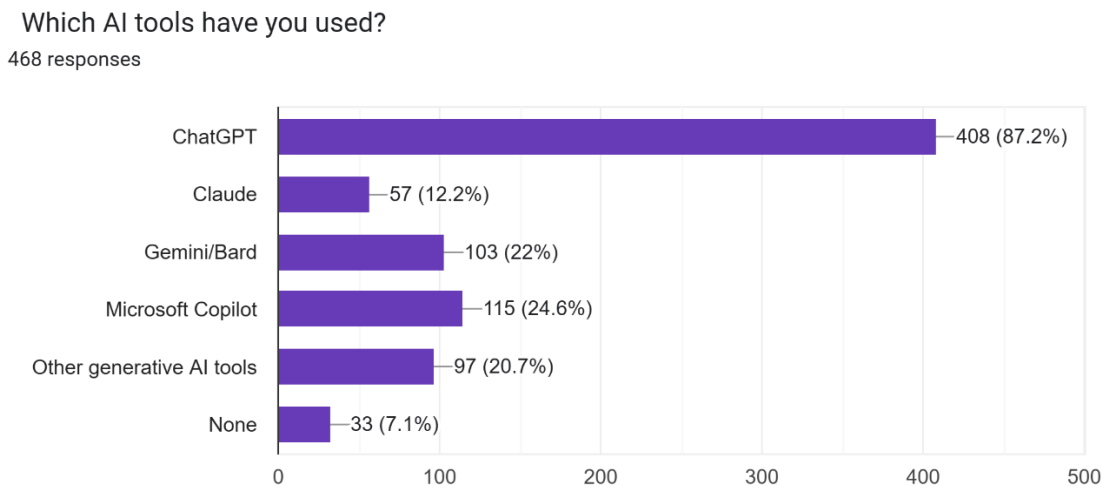

*Figure 4: Patterns of AI tool usage*

These findings clearly establish the need for a structured, scaffolded learning intervention focused on prompt engineering and responsible AI use. Participants were not approaching the training as complete novices, nor as experts, but as educators and researchers navigating early experimentation without a strong conceptual or ethical foundation. This reality set the stage for the webinar series, which was intentionally designed to move participants from casual and fragmented AI use toward intentional, ethical, and pedagogically grounded engagement. The webinars therefore responded directly to an identified capacity gap, particularly salient in Global South contexts, where access to advanced training, institutional guidance, and sustained professional development in AI remains uneven.

## 4. Webinar Series

This section presents the findings from the three webinar sessions, organised sequentially to reflect the programme's progressive structure. The results are reported as descriptive and interpretive accounts of participant engagement, discussion emphasis, and expressed orientations across sessions, rather than as evaluations of learning effectiveness or causal impact. Drawing on transcript-based indicators, facilitator observations, and computational summaries, the analysis illustrates patterns in discourse, sentiment, and topical focus that emerged during the professional development experience.

The webinar series was delivered online over three days using a live virtual platform. Each session lasted approximately 90 minutes and incorporated short lectures, live demonstrations, guided hands-on activities, and interactive discussion through chat and facilitator-led dialogue. The series was intentionally structured around a Progressive Learning Model, which supports the phased development of complex skills through sequenced and cumulative learning experiences. Progressive and scaffolded instructional approaches have been shown to strengthen engagement, retention, and higher-order reasoning in technology-mediated learning environments (Lu et al., 2025), making this model well suited for a short, intensive programme on generative AI. Each session therefore built deliberately on the previous one, moving from foundational concepts to applied prompt engineering and collaborative problem-solving. The blended delivery of lectures, demonstrations, guided practice, and reflective dialogue aligns with evidence that multimodal instruction enhances knowledge transfer in digital learning contexts (Haule et al., 2024).

Facilitation was led by Vanessa Willemse (E-Learning Division, South Africa), Dr. Benjamin Quarshie (GenAI-ERA, Ghana), and Matthew Nyaaba (GenAI-ERA; AI4STEM Education Center, USA). Participants included educators, postgraduate students, and early-career researchers from Ghana, Kenya, Lesotho, Nigeria, and South Africa. Their disciplinary grounding in teacher education and research positioned them to engage meaningfully with generative AI as an emerging professional literacy, while the transnational composition of the cohort reflected broader African ambitions to build AI capacity through context-responsive training.

Methodologically, the report draws on a qualitative, reflexivity-oriented programme evaluation design, commonly used by organisations and designers to analyse learning interventions and professional services (Vink & Koskela-Huotari, 2021; Lehtonen et al., 2015). Data sources included facilitator observations, reflective notes, and participant feedback, with the primary analytic corpus comprising YouTube transcripts of the three webinar sessions. Working with transcripts is increasingly recognised as methodologically robust when combined with computational techniques that support systematic organisation and transparency while retaining interpretive depth (Mettler, 2025). All transcripts were processed in Google Colab and analysed using Python, with generative AI tools (Gemini) supporting efficient data handling rather than automated interpretation.

Analytically, the study adopted a reflexivity-oriented stance that encourages continual examination of assumptions and interpretive decisions, consistent with the improvement-oriented goals of the GenAI-ERA initiative (Vink & Koskela-Huotari, 2021; Lehtonen et al., 2015). Transcript preprocessing followed established natural language processing procedures, including text cleaning, tokenisation, stop-word removal, and lemmatisation, in line with contemporary standards for qualitative text analysis (Mettler, 2025). These procedures were employed to structure the data systematically rather than to generate predictive or inferential models.

Four analytic layers informed the session-by-session analysis. Progressive learning indicators tracked the frequency and distribution of selected learning-related terms across transcript segments to examine shifts in discussion emphasis and expressed confidence over time, rather than to measure cognitive development. Reflexivity indicators examined the occurrence of selected self-referential linguistic markers (for example, "I think") as descriptive signals of personal positioning and judgement, not as measures of reflective depth. Thematic analysis employed predefined pedagogical and technological themes operationalised through clustered keywords to identify dominant discussion areas across sessions. Sentiment analysis, using the VADER algorithm, provided an additional descriptive lens on affective orientation, classifying transcript segments as predominantly positive, neutral, or negative (Mihai, 2024). Correlation tests were used exploratorily to examine relationships between thematic emphasis and sentiment trajectories, without implying causal association.

Findings were visualised through line plots, bar charts, heatmaps, and network diagrams to illustrate thematic evolution, sentiment distribution, and reflexive language patterns across sessions. Collectively, this multi-layered yet descriptive analytic strategy supports a coherent account of how participants engaged with prompt engineering and AI literacy across the three-day programme. The analyses are therefore presented as indicative patterns of participation and discourse dynamics, offering insight into how educators and researchers navigated AI-related concepts within a professional development context, rather than as definitive measures of learning outcomes or instructional impact.

### 4.1 Session 1: Prompt Foundations

*Session 1* established the conceptual foundations for prompt engineering. The facilitator introduced a structured framework that is *intent → context → instruction → constraint → output* to guide effective prompt construction. Participants engaged in practical reformulation exercises and live demonstrations using *Brisk* and *SciSpace*. Discussions highlighted how linguistic and cultural specificity, especially within African contexts, shapes AI interpretation. These interactions underscored prompt writing as both technical and communicative literacy essential for educators in AI-mediated learning environments. The analysis for this session is thematically presented along discussions such as *emerging central themes, sentiment orientations of participants and insights and future directions*.

*4.1.1. What is transcribed*

The dominant theme, AI in Education, reflected sustained interest in how AI tools are reshaping teaching, learning, and assessment, with high-frequency keywords (see Figure 5) signalling participants' strong engagement with AI's transformative potential. The analysis revealed that educators are increasingly conceptualising AI as a catalyst for pedagogical innovation rather than merely a technological add-on. Participants highlighted its capacity to personalise learning pathways, enhance instructional design, and improve teacher productivity perspectives that align with emerging scholarship emphasising AI's role in advancing learner-centred and data-informed pedagogies (UNESCO, 2023; Qazi & Pachler, 2024). The prominence of this theme within the session underscores a broader professional recognition that AI is rapidly becoming integral to educational practice, prompting educators to critically explore its promises, implications, and practical value in both classroom and institutional contexts.

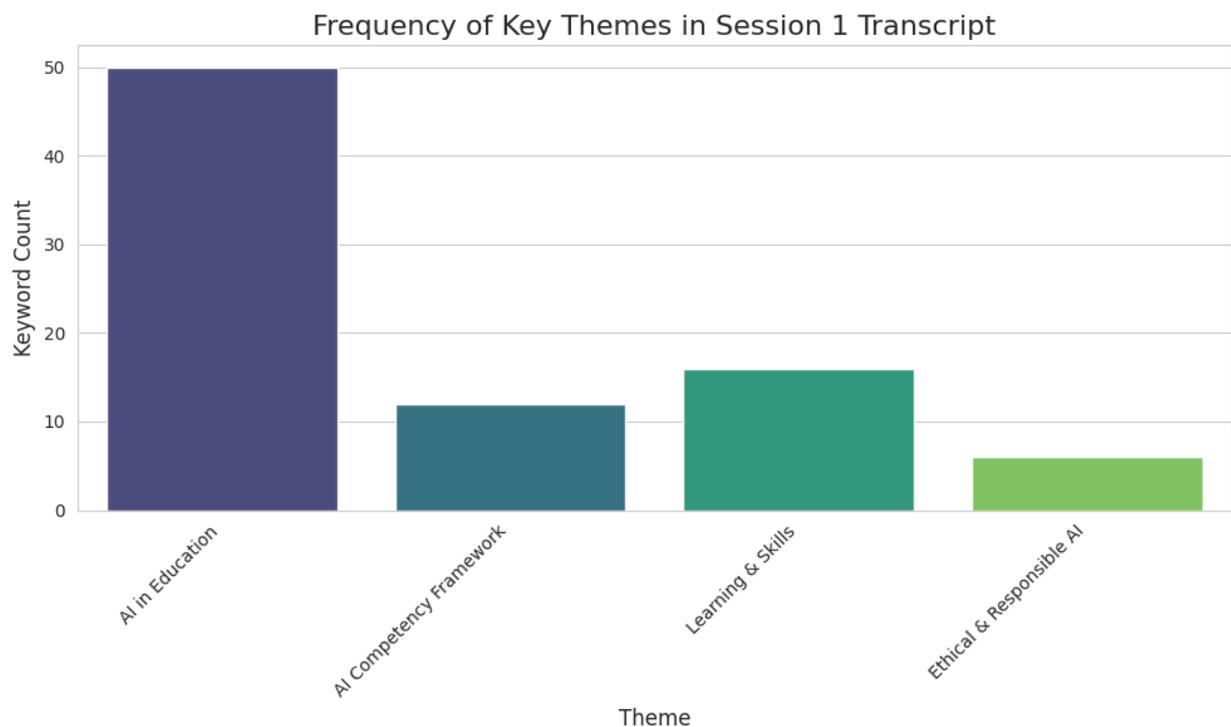

*Figure 5: High frequency themes from the session*

The next most highly emerging theme was Learning and Skills Development, which underscored educators' recognition of the need for new forms of professional competence to thrive in AI-mediated environments. Participants highlighted digital literacy, adaptive expertise, critical thinking, and continuous learning as essential competencies aligning with emerging frameworks on AI readiness in teacher education (Qazi & Pachler, 2024). The discussions demonstrated awareness that effective engagement with AI demands more than technical know-how; it requires conceptual understanding, ethical sensitivity, and the ability to exercise pedagogical judgement amid machine-generated outputs. Such emphasis on skills development is consistent with broader debates calling for teacher education programmes to embed AI literacy and data-informed pedagogy into their core curricula, particularly within African contexts where digital transformation remains uneven (Qazi & Pachler, 2024; Sperling et al., 2024).

A related but distinct dimension concerned the AI Competency Framework, which participants referenced as an authoritative structuring device for interpreting the capabilities and responsibilities required of educators in an AI-augmented landscape. The UNESCO AI Competency Framework for Teachers served as an implicit conceptual guide, offering shared language and direction for understanding responsible AI use (UNESCO, 2023). Its presence in the conversations, though less dominant than the other themes, indicates educators' desire for clarity, coherence, and standardisation as they navigate rapidly evolving technologies. This aligns with global trends where competency frameworks function as stabilising mechanisms that help educational institutions make informed decisions about curriculum, training, ethics, and infrastructure (Sperling et al., 2024). Participants' interest in competency standards suggests an appetite for structured, scaffolded professional development opportunities that demystify AI tools and enable purposeful classroom integration.

*4.1.2 Sentiment Orientations of Participants*

Across these themes, the sentiment orientations expressed by participants were largely optimistic. Educators articulated confidence in AI's potential to enhance teaching effectiveness, improve learning outcomes, and streamline administrative tasks. Demonstrations of AI tools appeared to reinforce a sense of forward momentum and professional empowerment. Our findings on optimism tempered by a clear desire for structure, clarity, and ethical-competency benchmarks reflect broader observations in AI-in-education scholarship, which note that while educators are eager to adopt AI, many lack formal conceptual frameworks and institutional support to guide responsible implementation (Sperling et al., 2024; Qazi & Pachler, 2024).

An emerging layer of ethical reflexivity was also evident, even though discussions on ethics were not dominant. Participants demonstrated awareness of issues relating to cultural representation, linguistic diversity, algorithmic bias, and responsible usage, concerns increasingly emphasised in African scholarship on digital technologies (Quarshie et al., 2025). Their reflections suggest that educators are beginning to interrogate the sociocultural implications of deploying AI tools that are often trained on datasets with limited African representation, a point central to ongoing conversations about digital justice and inclusion.

*4.1.3 Insights and Future Directions*

These collective insights indicate clear future directions for capacity-building initiatives. First, there is a need to deepen engagement with Ethical and Responsible AI, particularly in areas such as data governance, bias mitigation, transparency, and safeguarding learners. Structured ethical literacy training would strengthen educators' confidence in making informed decisions. Second, the application of AI Competency Frameworks should be operationalised through hands-on workshops that demonstrate how the standards can guide lesson planning, assessment design, and educational research. Such practical engagements will support meaningful, contextually relevant adoption. Finally, building on participants' enthusiasm, professional development programmes should expand skills development trajectories, including advanced prompting strategies, multimodal AI applications, and collaborative problem-solving using AI systems.

In all, the findings from this session highlight a teaching community that is curious, motivated, and increasingly reflective about the evolving role of AI in education. The first session provided a clear conceptual and practical entry point into prompt engineering, demonstrating strong readiness among educators to engage AI critically and constructively. The prominence of themes related to AI in Education and skills development underscores participants' appetite for meaningful integration, while the relatively modest emphasis on ethical considerations signals an important area for deeper inquiry in subsequent engagements. With targeted support, structured guidance, and ethically grounded capacity-building, educators can harness AI's potential while safeguarding the integrity, inclusiveness, and cultural responsiveness of teaching and learning in Ghana and beyond, thereby establishing a credible foundation for continued professional learning in AI-supported educational practice (Quarshie et al., 2025).

**4.2 Session 2 - Effective Prompts**

Session 2 advanced the foundational conversations from Session 1 by moving decisively from theoretical reflection to the responsible application of AI in educational practice (see Appendix B). The session foregrounded the concept of *prompt ownership* the ethical and intellectual accountability educators must exercise over AI-generated outputs while also addressing early participant questions about whether AI could legitimately write full research papers and how such work should be cited. The discussion revealed a discourse shaped less by abstract meta-theorising and more by pragmatic pedagogical concerns, as educators grappled with the realities of teaching, learning, and student engagement within an increasingly AI-mediated instructional landscape. Through meticulous preprocessing of the session transcript, a refined analytic corpus emerged, enabling a rigorous examination of the thematic and sentiment patterns underlying how educators articulate their instructional priorities and negotiate the opportunities and tensions accompanying emergent technologies.

*4.2.1. What transpired:*

The session unfolded through a structured progression of practical demonstrations, ethical deliberations, and collaborative critique. Facilitator 2 clarified early that while AI can enhance intellectual productivity, it cannot assume authorship in academic work; doing so would violate scholarly integrity and undermine the development of independent reasoning. This framing set the tone for the rest of the session, where participants worked in small groups to interrogate real AI-generated prompts that exhibited biases, inaccuracies, or conceptual gaps. Through iterative refinement, they collectively explored how intentional adjustments could transform flawed prompts into more accurate, context-sensitive queries. These activities illuminated a recurring ethical insight: effective prompting is inseparable from deliberate human judgement. The well-known principle "*garbage in, garbage out*" resonated strongly across discussions, underscoring that the quality of AI output remains fundamentally dependent on the user's epistemic responsibility, ethical positioning, and pedagogical intent.

The transcript evidence further shows that, despite the ethical framing, participants' engagement was primarily oriented toward practical pedagogical problem-solving rather than abstract theorisation. Progressive learning indicators (see Figure 6) revealed consistent reference to foundational concepts such as learning, knowledge, and development, but few explicit meta-cognitive markers such as insight, reflection, or transformation. This pattern aligns with Ghanaian teacher professional development traditions, which tend to favour solution-driven discourse, concrete instructional examples, and practical demonstrations over extended reflective commentary. It also mirrors global patterns in early-stage AI adoption, where educators often prioritise operational mastery before engaging in deeper epistemic critique (UNESCO, 2023).

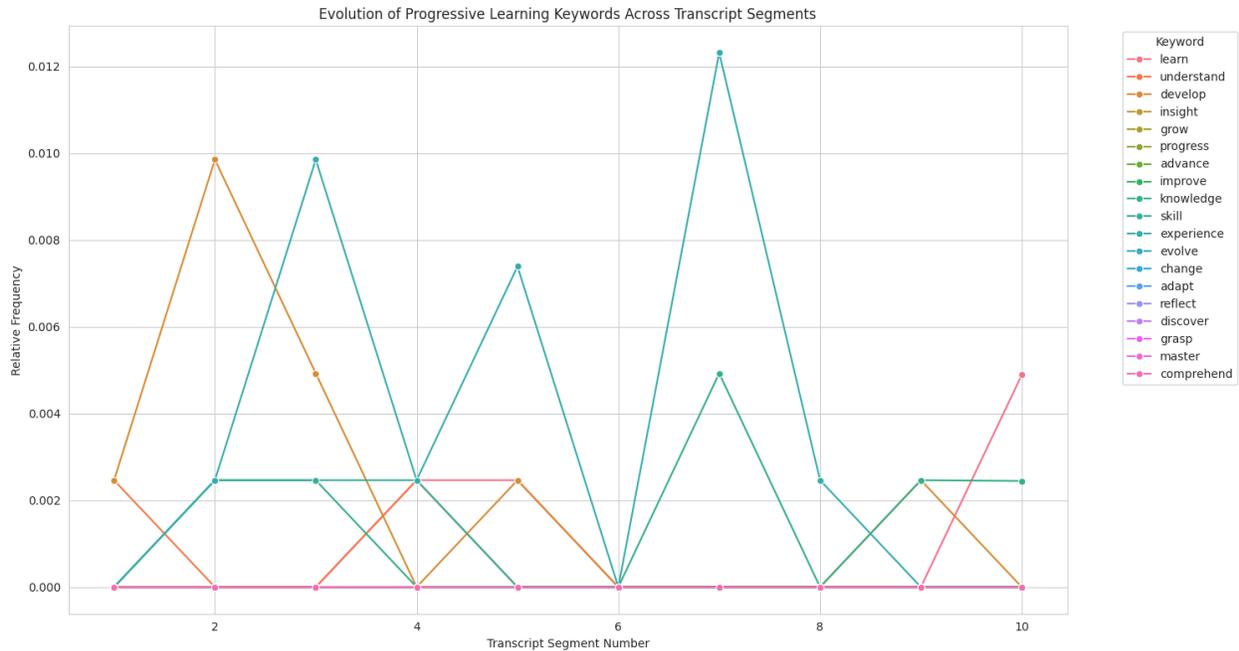

*Figure 6: Progressive learning Indicators emerging from the session 2*

Reflexivity patterns substantiate this interpretation (as discussed further under *4.3.1*) with its visual analysis of self-referential expressions (see Figure 11) showed variations of very high and low occurrences of phrases such as *I think*, *I feel*, or *we reflected*. Rather than indicating disengagement, this reflects the procedural orientation of the session, during which teachers focused on diagnosing challenges, comparing instructional techniques, and improving prompt structures. This behaviour parallels Session 1, which revealed that Ghanaian educators similar to their counterparts in many African contexts often frame technological learning through pragmatic, efficiency-oriented dialogues rather than personal epistemic positioning, a trend noted in broader African scholarship on educational technology integration (Abedi et al., 2023).

*4.2.2 Student engagement*

The thematic distribution reinforces this view. From the onset, *student engagement* emerged as the most dominant theme across transcript segments (see Figure 7), indicating that AI discussions were consistently anchored in the core pedagogical challenge of motivating learners within increasingly complex classroom environments. Participants repeatedly foregrounded the need for AI to support personalised learning, active participation, and more responsive instruction. Teaching methods also appeared prominently, demonstrating that educators were actively grappling with how AI might reshape lesson delivery, questioning strategies, and assessment practices. Technology, assessment, and collaboration surfaced intermittently, suggesting that while educators recognise AI's multidimensional implications, their immediate concerns remain situated within the heart of everyday instructional practice. However, collaboration_communication emerged strongly demonstrating acknowledgement of support systems which is key for progressive learning, first on the part of educators to onboard AI literacy and to support learners to develop needed competencies.

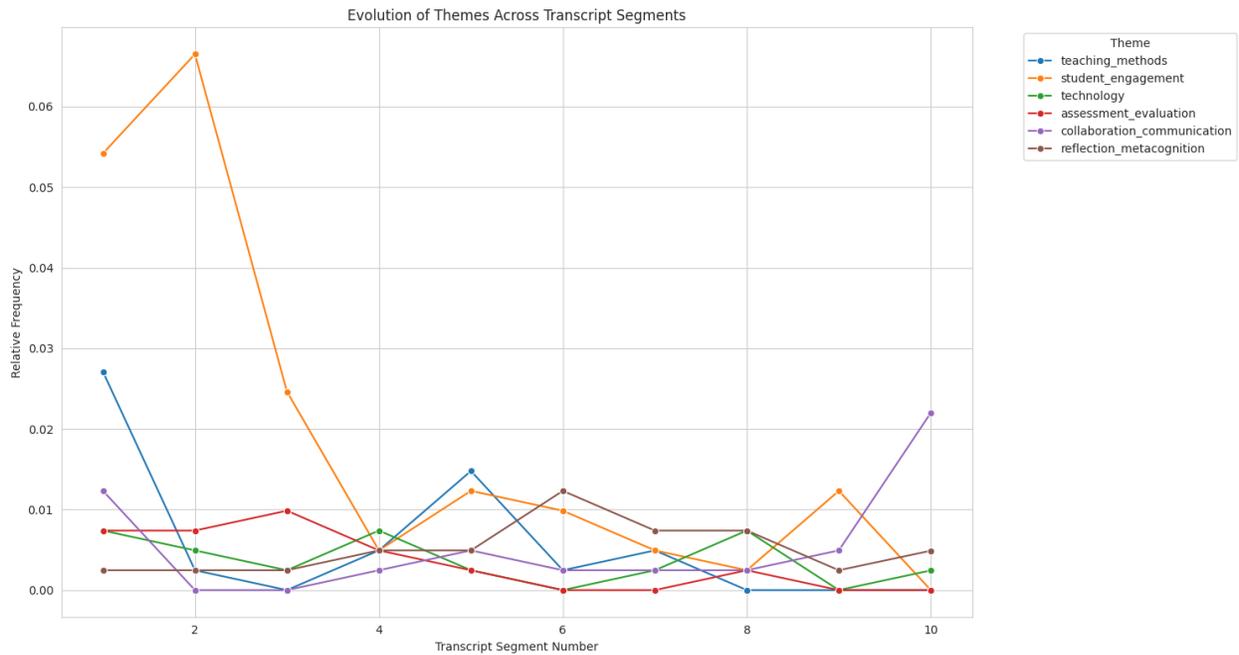

*Figure 7: Distribution of emerging themes across section 2*

### 4.2.3 Pedagogical challenges

This pattern signals an analytical yet optimistic orientation in that participants approached pedagogical challenges with seriousness but remained open, curious, and generally hopeful about the possibilities offered by AI. Generally as presented in *Figure 8*, there was an interesting distribution of sentiments across the section. Whereas *negative* and *positive* sentiments about using AI tools were expressed in minimal frequency, strong *neutral* and *compound* sentiments were recorded across the section. However, correlationally, slight reductions in positivity coincided with critical dialogues about teaching methods, assessment burdens, or technological constraints, an expected shift that aligns with global research indicating that deeper diagnostic deliberations naturally assume more sober tones. Correlation analysis (*see Figure 9)* strengthens this argument. Themes associated with problem-solving student engagement, teaching methods, technology correlated with higher neutral sentiment and marginally lower positivity, indicating thoughtful, evidence-based critique. Conversely, themes related to collaboration and reflection correlated with more positive sentiment, suggesting that when educators shared experiences or built ideas collectively, discussions took on a more affirming tone.

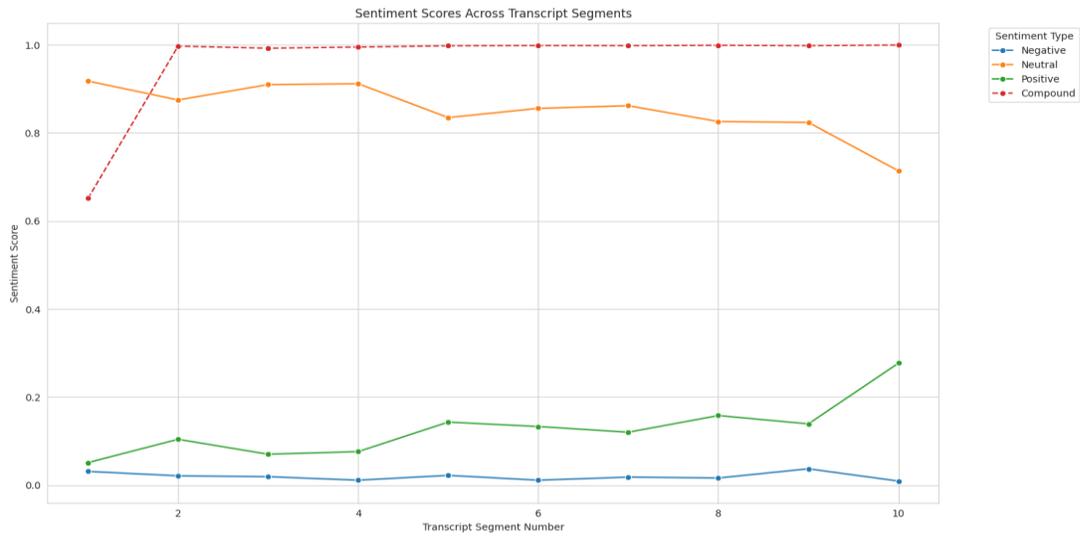

*Figure 8: Sentiment analysis across the section*

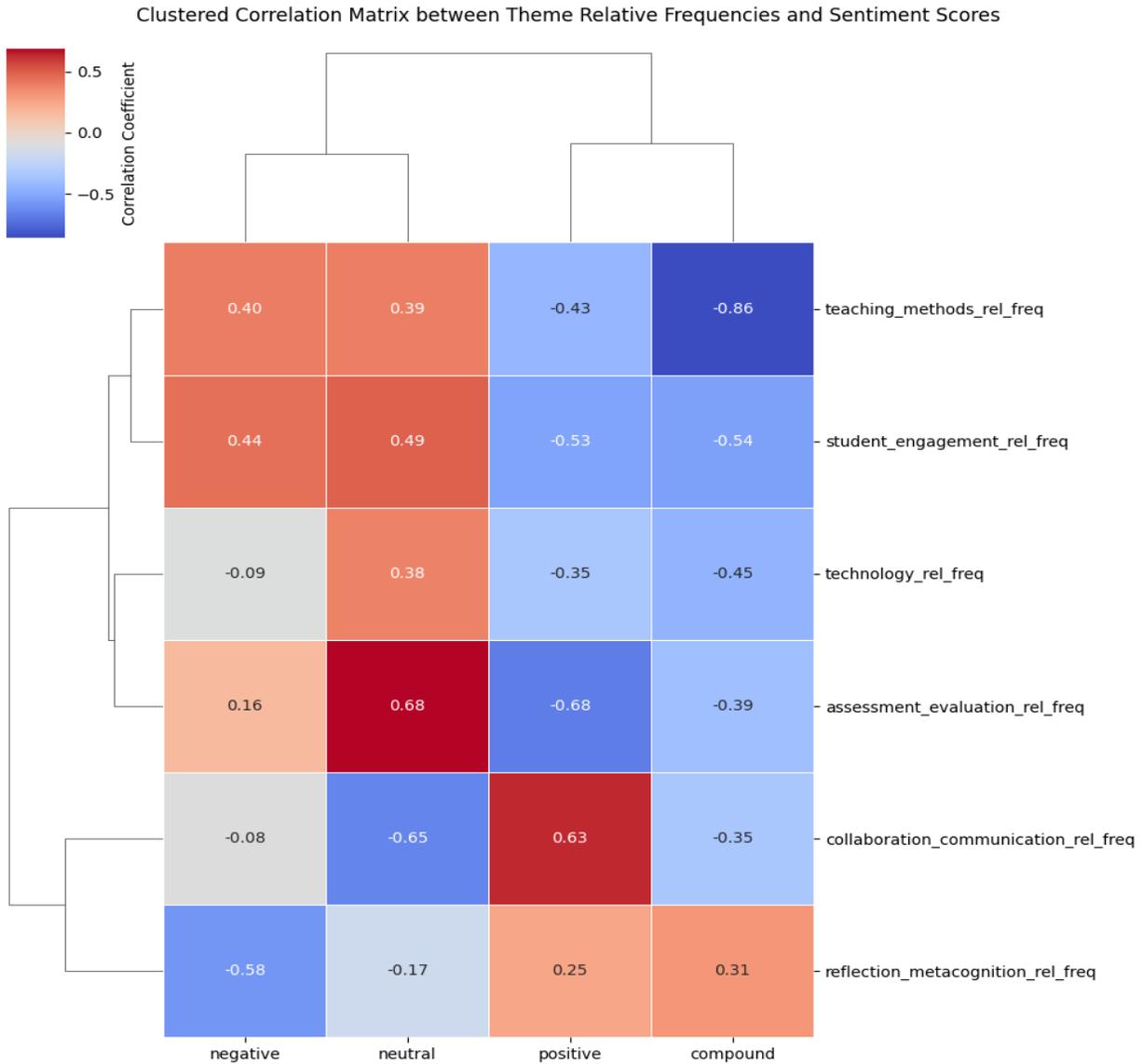

*Figure 9: Clustered heatmap showing correlation matrix between theme frequency and Sentiment Score*

*4.2.4 Professional community negotiating*

The discourse of Session 2 reflects a professional community negotiating the intersection of technological possibility and pedagogical responsibility. Educators demonstrated expanding conceptual clarity, growing technical confidence, and a willingness to interrogate both the opportunities and constraints of AI-mediated instruction. While explicit meta-cognitive reflection remained limited, the session revealed meaningful progression in professional judgement and pedagogical adaptability. This interpretive perspective is consistent with Session 1 yet expands it by showing how educators internalise and operationalise AI concepts when confronted with authentic instructional dilemmas. This pattern aligns with broader models of professional learning, where increasing conceptual mastery fosters deeper reflexive capacity when practitioners engage with authentic dilemmas (Vink & Koskela-Huotari, 2021; Lehtonen et al., 2015).

In essence, Session 2 extended the groundwork established in Session 1 by demonstrating how educators situate AI within the practical realities of teaching and learning, revealing a learning community that is analytically engaged, pedagogically anchored, and cautiously optimistic. The strong thematic emphasis on student engagement and instructional improvement highlights enduring Global South educational priorities, while the ethical framing around AI usage signals growing alignment with global academic integrity standards. The findings point to a professional cohort poised for deeper reflective exploration in subsequent sessions particularly on culturally responsive prompting, linguistic diversity, and ethical AI integration within Global South's evolving educational landscape.

**4.3 Session 3 – Practical Prompt Strategies and Personality Training**

Session 3 marked the transition from conceptual grounding (Session 1) and analytical engagement (Session 2) to advanced, application-oriented prompting strategies. The session introduced participants to zero-shot, few-shot, and chain-of-thought prompting, as well as the emerging use of personality-based prompt framing. Drawing on the Big Five Personality Framework, the facilitator demonstrated how prompt construction can intentionally shape AI tone, empathy, and stylistic orientation, an approach increasingly recognised in contemporary AI–human interaction research. This pedagogical shift mirrors global developments that emphasise not only technical proficiency but also socio-emotional calibration when working with generative AI systems (OECD, 2024).

The progressive learning analysis confirms that Session 3 was anchored in practical experimentation rather than theoretical exposition. Keywords such as *experience*, *learn*, and *understand* showed the highest frequencies (see *Figure 10*), underscoring the hands-on, application-driven nature of the session. However, much like earlier sessions, deeper metacognitive terms remained sparse, indicating that while participants actively engaged with new prompting strategies, explicit reflection on cognitive growth remained limited. This continuity with Session 1 and Session 2 suggests that the cohort's learning culture is practically inclined valuing experimentation and demonstration over meta-commentary.

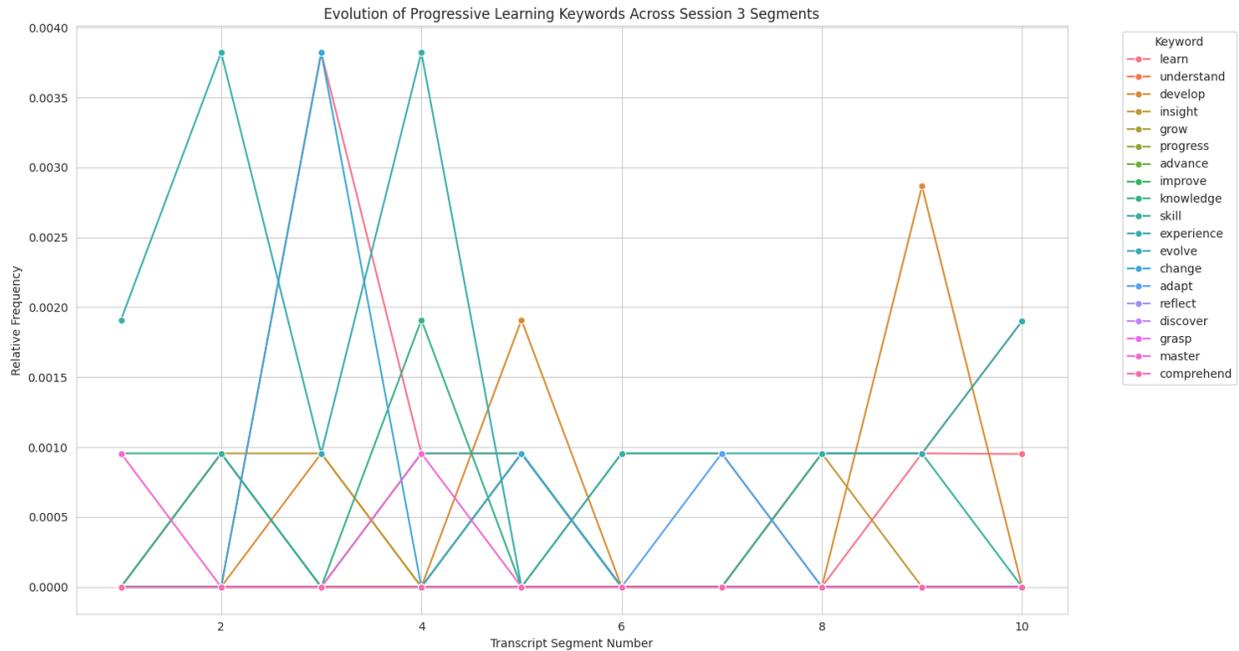

*Figure 10: Relative Frequency of Progressive Learning in session 3*

### 4.3.1 Cognitive Assertion and Confidence

Reflexivity patterns offer a notable departure from previous sessions. Session 3 saw a dramatic rise in the use of *"I think"* (62 occurrences), far exceeding the frequencies in Sessions 1 and 2 (see Figure 9). This indicates increased cognitive assertion and confidence as participants grappled with more complex prompting tasks. Other self-referential markers, such as *"myself"*, also appeared more frequently, demonstrating a gradual emergence of personal voice within the learning environment. Despite this individual reflexivity, collective forms of reflection such as *"we discussed"* remained almost absent, reinforcing a consistent trend across all sessions. The data points to a reflective style that is individually expressive but not yet collaboratively articulated.

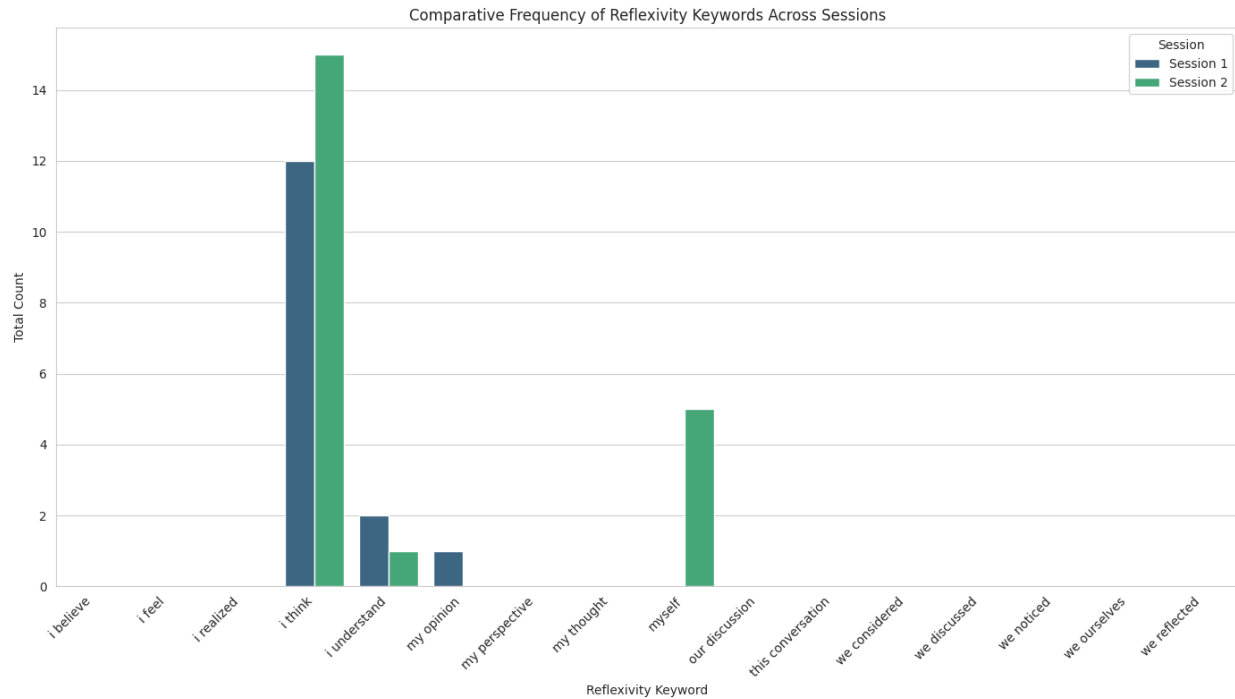

*Figure 11: Comparative Reflexivity Keywords across sections 1&2*

### 4.3.2 AI personas, and testing chain-of-thought reasoning

Thematic analysis reveals a shift in conceptual emphasis. In contrast to Session 2, which prioritised *student engagement*, Session 3 projects *technology* as the most prominent theme at the end of the section (see Figure 12). This is not surprising given that participants were actively manipulating prompts, analysing AI personas, and testing chain-of-thought reasoning. Discussions of *student engagement* and *teaching methods* remained visible but were more supplementary, indicating that participants were now grappling with the technical underpinnings that enable more sophisticated pedagogical applications, reflective of progressive learning. As in Sessions 1 and 2, *reflection/metacognition* remained minimally represented, highlighting a series-wide opportunity to strengthen explicit reflective practice as AI use becomes more deeply embedded in educational work. Comparatively, there is more to do with *teaching_methods* as it recorded very low frequency visibility at the end of the section. To strengthen AI literacy, educators must be intentional about opening conversations about AI-based pedagogical issues while acquiring the needed skills to drive progressive learning.

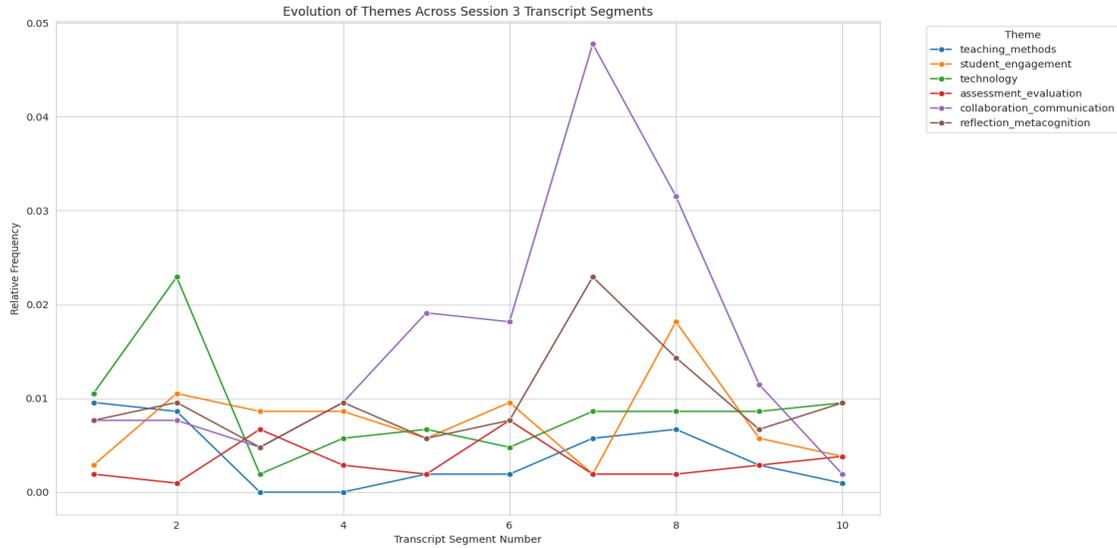

*Figure 12: Emerging themes from section 3*

### 4.3.3 Advanced Prompt and Anxiety

Sentiment analysis shows remarkable consistency with the earlier sessions. Neutral sentiment dominated (84%), accompanied by a strong positive tone (13.6%) and minimal negativity (2.4%) (see Figure 13). This emotional profile suggests that the introduction of advanced techniques did not produce anxiety or frustration; rather, the learning environment remained constructive and highly supportive. This aligns with findings from Sessions 1 and 2, which showed that Ghanaian educators increasingly engage AI-related professional learning with optimism, curiosity, and epistemic openness.

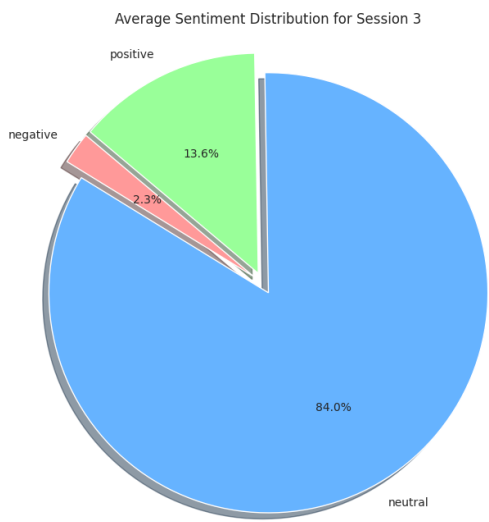

*Figure 13: Distribution of Sentiments in Session 3*

*4.3.4 Persona-guided prompting to humanise AI interactions*

Through the discussions, participants explored the pedagogical and ethical implications of personality-based prompting. Activities in which participants assigned AI roles such as supportive tutor, sceptical reviewer, or culturally empathetic advisor brought to the fore the potential of persona-guided prompting to humanise AI interactions and enhance culturally responsive pedagogy. This is particularly relevant in African educational contexts, where linguistic diversity, cultural nuance, and differentiated learner experiences demand adaptable and sensitive instructional tools. As participants observed, however, personality simulation introduces risks of implicit bias, over-personalisation, or misrepresentation reinforcing the need for human oversight, an issue first raised in Session 1's discussions on ethical AI use. The session's final reflections emphasised that while personality-based prompting can enrich educational engagement, its responsible use requires both technical fluency and ethical awareness.

In synthesis, Session 3 reflects a leverage on the progressive learning indicators by deepening the pedagogical sophistication of the series through extending participants' capacities to manipulate AI reasoning patterns and emotional tone through advanced prompting strategies. The rise in reflexive expression indicates growing confidence and intellectual agency among educators, even as collective reflection remains underdeveloped. The strong emphasis on technology, aligned with sustained positive sentiment, points to a community increasingly prepared to integrate AI meaningfully provided that emerging ethical concerns are addressed in continued professional development. Together with Sessions 1 and 2, this session underscores the need for a balanced approach that combines technical mastery, reflective depth, and cultural responsiveness as AI becomes a more integral component of educational practice in Ghana and beyond.

## 5. Cross-Session Reflections

This cross-session analysis synthesises participant reflections from three webinar sessions with findings derived from progressive learning, reflexivity, thematic, and sentiment analyses. Together, these data points provide a holistic view of how educators engaged with the topics, how discussions evolved, and the underlying dynamics shaping their views on AI integration in education.

### 5.1 Prompt Quality Consistently Framed by Participants as Influencing

Across all sessions, participants repeatedly affirmed that the usefulness and reliability of AI outputs depend largely on the clarity, specificity, and contextual depth embedded in the prompts. This insight was strongly reinforced by the quantitative patterns observed across the sessions. Discussions categorised under the broad "technology" theme increased progressively, signalling sustained engagement with the practical use of AI tools. Even though the thematic categories did not isolate "prompting" as a standalone theme, the rising emphasis on technology-centred dialogue suggests that participants were increasingly concerned with how to refine their interactions with AI systems to obtain more meaningful results. The sentiment analyses further reinforce this interpretation. Across all three sessions, educators maintained predominantly positive and neutral tones when engaging with technically demanding topics. This indicates that discussions around prompt construction, though sometimes challenging, were framed as constructive and collaborative problem-solving exercises (Ivanov & Song, 2024). The combined qualitative and quantitative insights reveal that participants not only recognised the importance of effective prompting but were also actively refining this skill as part of their developing AI literacy.

### 5.2 Ethical Literacy Equals AI Literacy

Participants consistently emphasised that responsible engagement with AI requires deep ethical awareness, including issues of bias, data provenance, and scholarly integrity. Although ethical concerns did not appear as dominant quantitative themes due to the limitations of the keyword-based approach, the persistent if modest presence of reflection-oriented discourse across sessions suggests periodic moments of critical engagement with ethical implications. The correlation between reflective themes and positive sentiment indicates that participants approached discussions on ethics constructively rather than defensively or with apprehension. While explicit markers of deep ethical deliberation were relatively low, the sessions nonetheless demonstrated a growing consciousness of the need for critical oversight when integrating AI into teaching, assessment, and research practices. The findings suggest that educators view ethical literacy not as an optional add-on but as an essential component of AI competence, one that must be intentionally cultivated in professional learning environments (Dilek et al., 2025). It is interesting to note that ethical also appeared episodically, however, its importance is inferred through recurrent but low-frequency markers thereby emphasising orientation rather than depth.

### 5.3 Human Creativity Remains Central

Another clear cross-session insight suggests that educators view AI as an augmenting tool rather than a substitute for human reasoning, creativity, or moral judgment. Participants highlighted the importance of maintaining pedagogical agency, contextual awareness, and value-based decision-making in AI-supported environments. This view is reflected subtly but consistently in the quantitative findings. The progressive learning analysis indicates that while foundational learning terms such as "learn," "understand," and "experience" were present, deeper cognitive terms associated with insight, reflection, and conceptual mastery appeared infrequently. This pattern suggests that while participants engaged actively with foundational aspects of AI-supported learning, the uniquely human features of teaching such as creativity, discernment, and contextual decision-making were understood intuitively rather than verbalised in these particular lexical forms (Nyaaba et al., 2025). Additionally, themes such as "student engagement" and "teaching methods" remained strong across sessions, indicating that educators anchored their discussions on pedagogical concerns rather than technological determinism. This confirms that participants interpret AI not as a replacement for the educator, but as a tool whose value lies in enhancing human-centred teaching and learning (Nyaaba & Zhai, 2025).

## 5.4 Practical Needs in Low-Resource Contexts

Participants also emphasised contextual challenges associated with AI adoption in low-resource educational settings. These include affordability of AI platforms, the need for locally relevant training materials, and sustained institutional support for digital capacity-building. The thematic analyses reflect increasing attention to the "technology" category, suggesting heightened interest in practical applications and infrastructural considerations (Nyaaba et al., 2024).

Similarly, positive and neutral sentiment across sessions indicates that participants approached these challenges with a forward-looking mindset rather than frustration or resignation (Lee et al., 2025). Discussions related to collaboration and communication point to the need for shared resources, peer learning networks, and institutional policies that support equitable access to AI tools. Although the quantitative data do not directly capture concepts such as affordability or localisation, the strong emphasis on practical, classroom-oriented themes suggests that participants framed AI adoption in terms of actionable strategies that must align with contextual realities.

## 5.5 Cross-Session Trends

Across the three sessions, the discussions were consistently pragmatic, constructive, and forward-looking. Findings reflect patterns in discourse and expressed confidence. Reflexivity analysis indicates a notable increase in the use of statements such as "I think," signalling that participants became progressively more confident in articulating personal perspectives. However, both reflexive keywords and reflection-based thematic categories remained generally low across sessions, indicating that deeper collective reflection and meta-cognitive engagement have not yet matured within the group. The shift toward technology-focused themes in the final session reflects evolving interest in practical implementation, experimentation, and tool-based problem-solving. At the same time, educators remained attentive to ethical considerations and the irreplaceable role of human judgment in AI-mediated contexts, reflecting a general concern of most African educators higher institutions (Quarshie et al., 2025). In all, the integration of qualitative insights and quantitative analyses paints a picture of an educator community that is curious, adaptive, and eager to build competence yet still in the early stages of developing deeper reflective and ethical engagements with AI. The sessions ultimately serve as dynamic platforms for exploring both the possibilities and constraints of AI in education. While participants actively exchanged strategies and perspectives, the findings also point to an opportunity to strengthen structured reflective practice, ethical engagement, and institutional support as AI adoption becomes more central to educational transformation.

## 6. Conclusion and Recommendations

The GenAI-ERA three-day Prompt Engineering Training Series provided a robust platform for building educators' and researchers' capacity to engage meaningfully with generative AI. Across the three sessions, participants demonstrated growing confidence, conceptual clarity, and practical competence in applying prompt engineering to teaching, assessment, and research. The combined qualitative and quantitative analyses reveal an educator community that is increasingly curious, reflective, and willing to experiment with AI-enhanced pedagogical strategies, even within resource-constrained environments.

Overall, the discourse across sessions was positive, pragmatic, and action-oriented. Thematic patterns showed a consistent emphasis on *student engagement* and *teaching methods*, while the prominence of *technology* increased steadily signalling educators' readiness to embed AI tools into everyday practice. Reflexivity patterns also evolved, with a marked rise in statements such as "I think," suggesting increased individual confidence in articulating perspectives on AI use. However, explicit collective reflection and deeper metacognitive discourse remained limited, pointing to an opportunity for future training to cultivate more structured reflective practices.

Despite the technological focus, participants repeatedly emphasised that human creativity, ethical reasoning, and contextual intelligence remain central to AI integration in African education. The sessions underscored that prompt engineering is not merely a technical skill but a composite literacy that combines precision, ethical grounding, contextual awareness, and pedagogical judgment. The overwhelmingly positive emotional tone across discussions further reflects a learning environment in which participants approached challenges constructively and collaboratively.

At the same time, the analyses highlight contextual concerns that must be addressed to ensure equitable access to AI skills. Participants noted the need for affordable AI tools, locally relevant training materials, and institutional systems that can sustain continuous professional development. The findings therefore suggest that meaningful AI adoption in African education depends not only on individual competence but also on supportive policy frameworks, infrastructure, and institutional capacity. These insights affirm that African educators and researchers are positioned not merely as users of global AI technologies but as active contributors capable of shaping ethical, culturally responsive, and contextually grounded AI practices.

## 6.1 Recommendations

Universities, colleges of education, and training institutes should embed prompt-engineering and AI-literacy modules within digital literacy, research methods, and professional pedagogy courses. Embedding these competencies institutionally ensures that AI literacy becomes a foundational element of teacher and researcher development rather than an optional add-on. National education ministries, regulatory bodies, and professional councils should develop clear guidelines for ethical AI use, authorship, academic integrity, and responsible citation practices. These policies are essential for safeguarding scholarly standards and ensuring culturally relevant, bias-aware AI integration.

To bridge disparities between urban and rural institutions, AI training materials should be adapted into low-bandwidth, offline, and multilingual formats. This includes printable toolkits, voice-based modules, and locally contextualised examples that reflect Ghanaian and African pedagogical realities. Institutions should establish communities of practice, mentorship structures, and peer-learning networks to support continuous capacity building. Regular refresher workshops, collaborative prompt-engineering labs, and inter-institutional partnerships will deepen long-term competence. Longitudinal studies should be undertaken to assess how prompt-engineering training influences teaching quality, research productivity, and learner outcomes over time. Such research will provide an evidence base for refining curricula, improving training models, and informing national policy. Future analyses of similar training series should incorporate more sophisticated NLP and discourse analysis techniques such as topic modelling, semantic embeddings, and speaker-specific tracking to capture deeper patterns of reflection, collaboration, and thematic evolution across sessions. Prioritising structured AI-literacy initiatives, coupled with contextual relevance and ethical grounding, could potentially enable African educators and researchers to transition from passive consumers of global AI tools to active contributors shaping the future of AI in education. The GenAI-ERA initiative demonstrates that with intentional design, collaborative engagement, and sustained support, AI capacity-building can thrive even in low-resource contexts, marking an important step toward a more equitable and innovative African digital future.

**Acknowledgments**

The authors thank all participants, institutional partners, and the GenAI-ERA team for technical support. Special appreciation goes to the facilitators namely Vanessa Willemse, Dr. Benjamin Quarshie, and Matthew Nyaaba, for their leadership and to the GenAI-ERA executive team for organizing and documenting the event.

**Conflict of Interest**

The authors declare no conflict of interest.

**Appendix A**

GenAI-ERA website: ( https://genaited.com/)

*GenAI-ERA LinkedIn: https://www.linkedin.com/company/genai-for-educators-and-researchers-in-africa-genai-era/posts/?feedView=all*

*GenAI-ERA partnership with UNESCO MOST-BRIDGE: https://bridges.earth/coalition-members*

**Appendix B**

# Prompt iteration strategies

- Prompt design can sometimes require a few iterations before you consistently get the response you're looking for.

| Use different phrasing | Switch to an analogous task: | Change the order of prompt content: |
|---|---|---|
| Version 1: How do I bake a pie? | Prompt: Which category does The Odyssey belong to: thriller sci-fi mythology biography | Version 1: [examples] [context] [input] |
| Version 2: Suggest a recipe for a pie. | | Version 2: [input] [examples] [context] |
| Version 3: What's a good pie recipe? | | Version 3: [examples] [input] [context] |

# Example of Few-Shot

**Prompt:**
Please choose the best explanation to the question:

Question: How is snow formed?

**Explanation1:** Snow is formed when water vapor in the air freezes into ice crystals in the atmosphere, which can combine and grow into snowflakes as they fall through the atmosphere and accumulate on the ground.

**Explanation2:** Water vapor freezes into ice crystals forming snow.
Answer:

**Response:**
**Explanation1** is the better explanation because it provides more detail on the process, including how ice crystals combine and grow into snowflakes as they fall through the atmosphere.

## Partial input completion

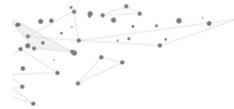

- Generative language models work like advanced autocomplete.
- You give partial content; the model predicts and generates the continuation.
- The output is what the model "thinks" should logically come next.
- Examples you provide guide the style, structure, and format of the response.
- Context you provide (topic, audience, purpose) shapes relevance and accuracy.
- Prompts often combine an **instruction** (what to do) and an **entity/input** (what to act on).

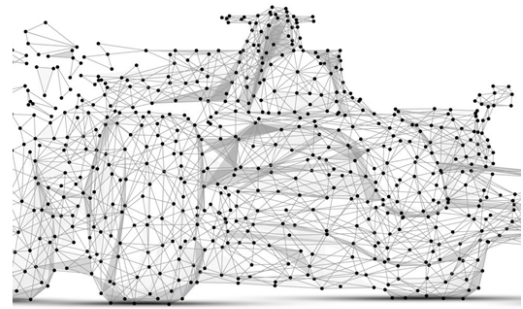

## General Text Prompt

**Baseline (weak):**
"Explain photosynthesis."

**Revision (better):**
"You are a high-school biology tutor. In ≤150 words, **explain photosynthesis** to a 10th-grader using **plain language** and **one everyday analogy**. End with **three study flashcards** in the format Q:/A:."

**Why better?** Role + audience + length + analogy + deliverable structure.

## Zero-shot vs few-shot prompts

You can include examples in the prompt that show the model what getting it right looks like.

Prompts that contain a few examples are called *few-shot* prompts, while prompts that provide no examples are called *zero-shot* prompts.

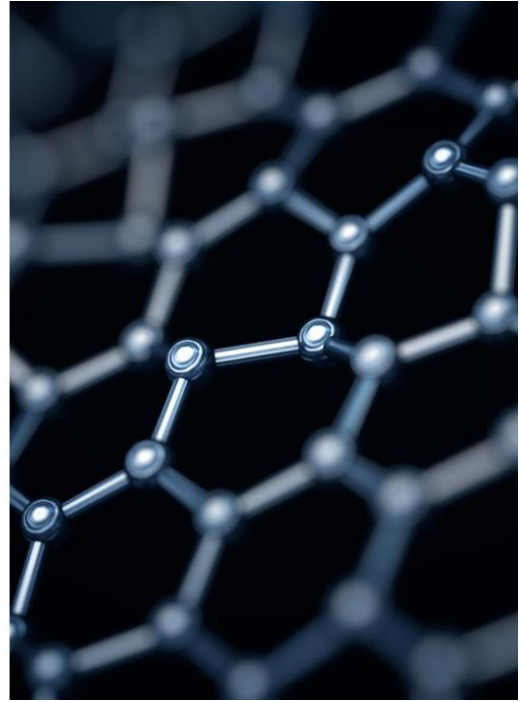

# Key Elements of an Effective Prompt Strategy

- **Goal:** One clear outcome (what "good" looks like).
- **Audience & Tone:** Formal vs. friendly; technical vs. plain-language.
- **Scope & Constraints:** Length, style, sources, timespan, jurisdiction.
- **Inputs & Context:** Only relevant info; avoid irrelevant noise.
- **Examples (Few-Shot):** Show correct format and level of detail.
- **Output Format:** JSON/table/bullets, plus any required fields.
- **Quality Check:** Ask for verification steps or a short rationale.
- **Safety & Ethics:** Prohibit harmful content; request respectful phrasing.

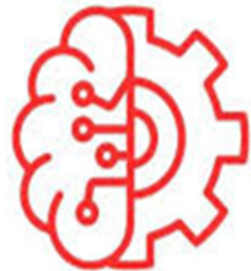